\documentclass[aps,preprintnumbers,11pt,amsmath,amssymb,nofootinbib]{revtex4}

\usepackage{braket}
\usepackage{float}
\usepackage{graphics,epsfig}
\usepackage{bm}
\usepackage{slashed}
\usepackage{graphicx}
\usepackage{amsmath}
\usepackage{amsfonts}
\usepackage{amssymb}
\usepackage{color}%
\usepackage{ulem}
\usepackage{booktabs}
\usepackage{dcolumn}
\usepackage[
            pdfstartview=FitH,
            bookmarksnumbered=true,
            bookmarksopen=true,
            colorlinks,
            linkcolor=blue,
            anchorcolor=green,
            citecolor=red
            ]{hyperref}
\usepackage{enumerate}
\providecommand{\U}[1]{\protect\rule{.1in}{.1in}}

\def\A0{A^{(0)}}

\begin{document}
\baselineskip=0.6 cm \title{Towards a Fisher-information description of complexity in de Sitter universe}

\author{Chong-Bin Chen$^{1,2}$}
\thanks{E-mail address: cchongb23@gmail.com}
\author{Fu-Wen Shu$^{1,2,3,4}$}
\thanks{E-mail address: shufuwen@ncu.edu.cn}
\affiliation{
$^{1}$Department of Physics, Nanchang University, Nanchang, 330031, China\\
$^{2}$Center for Relativistic Astrophysics and High Energy Physics, Nanchang University, Nanchang 330031, China\\
$^{3}$GCAP-CASPER, Physics Department, Baylor University, Waco, TX 76798-7316, USA \\
$^{4}$ \quad Center for Gravitation and Cosmology, College of Physical Science and Technology, Yangzhou University, Yangzhou 225009, China}

\vspace*{0.2cm}
\begin{abstract}
\baselineskip=0.6 cm
\begin{center}
{\bf Abstract}
\end{center}
Recent developments on holography and quantum information physics suggest that quantum information theory has come to play a~fundamental role in understanding quantum gravity. Cosmology, on the other hand, plays a~significant role in testing quantum gravity effects. How to apply this idea~to a~realistic universe is still unknown. Here, we show that some concepts in quantum information theory have cosmological descriptions. Particularly, we show that the complexity of a~tensor network can be regarded as a~Fisher information measure (FIM) of a~dS universe, followed by several observations: (i) the holographic entanglement entropy has a~tensor-network description and admits a~information-theoretical interpretation, (ii) on-shell action of dS spacetime has a~same description of FIM, (iii) complexity/action(CA) duality holds for dS spacetime. Our result is also valid for $f(R)$ gravity, whose FIM exhibits the same features of a~recent proposed $L^n$ norm complexity.
\end{abstract}

\maketitle
\newpage
\vspace*{0.2cm}

%
\section{Introduction}

A milestone in the exploration of the unification of general relativity and quantum mechanics was the work of Bekenstein and Hawking on the area~law of black hole entropy~\cite{Bekenstein:1973ur,Hawking:1974rv}. Inspired by this discovery, 't Hooft~\cite{tHooft:1993dmi} and Susskind~\cite{Susskind:1994vu} formulated the holographic principle, which suggests that the degrees of freedom of a~higher dimensional gravitational system can be characterized by those of a~lower dimensional quantum system. This principle is currently widely regarded as a~fundamental principle of quantum gravity, especially after Maldacena's discovery~\cite{Maldacena:1997re,Witten:1998} of AdS(Anti-de Sitter)/CFT(Conformal field theory) correspondence.

However, how these extra~degrees of freedom emerge from CFT is still a~mystery. A breakthrough came from the~recently proposed holographic entanglement entropy (HEE)~\cite{Ryu:2006bv}, which suggests deep connections between quantum gravity theory and quantum information theory~\cite{VanRaamsdonk:2010pw,Swingle:2009bg}. However, although these connections are generally believed to grasp a~significant character of the theory of quantum gravity, there is a lack of applications to the realistic universe. Most current achievements are valid only for AdS spacetimes, with very limited efforts to our realistic universe.

In this work, we try to make a~preliminary attempt to cross these gaps. We focus on the possible relations between the Friedmann-Robertson-Walker (FRW) universe (particularly the dS universe) and quantum information theory. We show that complexity of a~multi-scale entanglement renormalization ansatz (MERA)~\cite{Vidal:2008zz} tensor network can be thought of as FIM of a~dS spacetime. Our argument is based on the following three observations: First, we will show that for MERA tensor network, the~entanglement entropy of a~cut leg can be viewed as a~flow---an information-bit (qubit) flow transmitted by a~quantum circuit. It provides an information-theoretical picture of the MERA network. According to this picture, tensor network and spacetimes admit the same causal structure. This is consistent with the MERA/spacetime correspondence proposed in~\cite{Beny:2011vh}, where MERA is regarded as a~quantum circuit and the dS metric is derived. A similar perspective can be found in~\cite{Czech:2015qta,Czech:2015kbp,Asplund:2016koz,Czech:2016xec,deBoer:2015kda}, where MERA is viewed as a~discretization of kinematic space---the space of bulk geodesics, instead of the time slice of the original bulk, and the kinematic space of an AdS space is of dS geometry. Second, the on-shell action of dS can be identified as an~FIM, which is a~description of the measure of information. Third, we show that the on-shell Einstein-Hilbert (EH) action of dS spacetime exhibits the same features as complexity of MERA network, which can be viewed as a~dS-version complexity/action(CA) duality~\cite{Brown:2015bva,Brown:2015lvg,Chapman:2016hwi,Carmi:2016wjl}. Originally, the CA duality refers to the Wheeler-De Witt(WDW) patch under asymptotic AdS spacetime. In our dS-version CA duality, we do not need to constrain in the WDW patch. This follows from recent generalization, for instance, complexity of MERA in terms of Liouville action as shown in~\cite{Caputa:2017urj,Caputa:2017yrh,Czech:2017ryf,Caputa:2017}, and~dS spacetime as shown in~\cite{Bao:2017qmt}.

\section{MERA/de Sitter Correspondence}

\begin{figure}[H]
\centering
\includegraphics [scale=0.6]{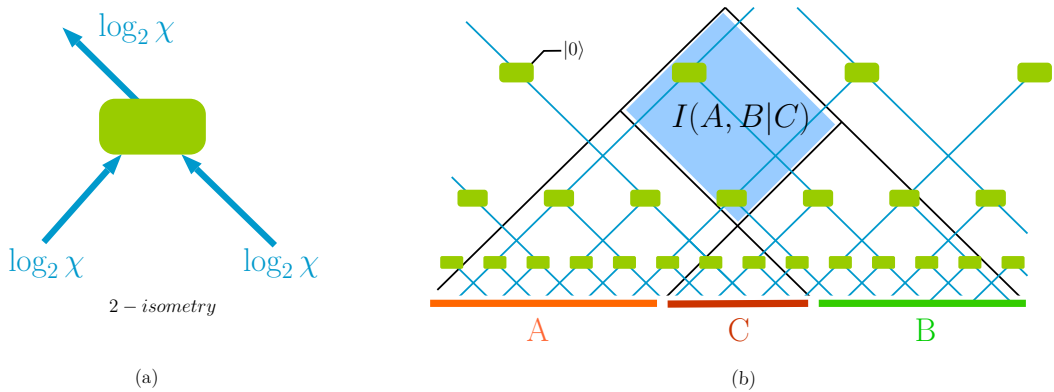}
\caption{(\textbf{a}) A $2-$isometry tensor element. (\textbf{b}) The multi-scale entanglement renormalization ansatz (MERA) network. We have ignored disentangler because the spacetime volume is only interpreted as conditional mutual information. $A$ and $B$ share the information $I(A,B|C)$~\cite{Czech:2015kbp}. These entangled degrees of freedom are transmitted by isometry in the blue region. }\label{TN}
\end{figure}

Given a~MERA network, without loss of generality, we assume it is a~2-isometry tensor network, which means each isometry in the network has two lower legs and one upper leg. Cutting one leg will gives $\log_{2}\chi$ entropy~\cite{Vidal:2008zz}, where $\chi$ is the bond dimension. The key point is that 2-isometry is a~coarse-graining operator mapping $\chi^{2}$-dimensional Hilbert space to $\chi$-dimensional one as illustrated in Figure~\ref{TN}a. This property suggests that $\log_{2}\chi$ can be regarded as flux of entanglement flow in each leg and causal relation between tensors can be viewed as causal structure of the emergent spacetime~\cite{Swingle:2009bg,Czech:2015kbp}. The entanglement entropy is given by counting the number of legs on the causal cut. Before discussing this MEAR/spacetime correspondence, let us first give a~quick snapshot of tensor network in terms of the kinematic space of AdS$_3$. According to~\cite{Czech:2015qta,Czech:2015kbp,Asplund:2016koz,Czech:2016xec,deBoer:2015kda}, MERA tensor network is best viewed as kinematic space of AdS$_3$ rather than the time slice of the original AdS$_3$. The kinematic space is defined by a~set of boundary-anchored geodesics. The measure of a~kinematic space is determined by~\cite{Czech:2015qta,Asplund:2016koz} $\mathcal{D}g \propto (\partial^2S(u,v)/\partial u\partial v)dudv$, which is the measure of dS$_2$.

In terms of this picture, the volume of a~causal diamond $D$ can be explained as conditional mutual information of two intervals as shown in Figure~\ref{TN}b: $I(A:B|C)=S(u-du,v)+S(u,v+dv)-S(u,v)-S(u-du,v+dv)=(\partial^2S(u,v)/\partial u\partial v)dudv.$ In other words, isometries in the region $D$ share the information of $A$ and $B$ so they contain the entanglement degrees of freedom between $A$ and $B$. The number of information bits in the region $D$, which is denoted as $\mathcal{N}$, is proportional to the volume integral over $D$,
\begin{equation}\label{number}
\mathcal{N} = \int_{D} d^2x \sqrt{-g}\Theta_{I},
\end{equation}
where $\Theta_{I}$ is a~constant 
and $g_{\mu\nu}$ is the dS$_2$ metric with radius $L$. The idea~that measuring the volume of a~region in a~manifold is replaced by counting the number of elements in this region, was first suggested by Riemann~\cite{Riemann}. This is also the main idea~of some quantum gravity models such as causal sets theory~\cite{Bombelli:1987aa} and dS/MERA correspondence~\cite{Beny:2011vh,Czech:2015qta,Czech:2015kbp,SinaiKunkolienkar:2016lgg,Bao:2017qmt}.

Note that the number of information bits in the diamond $D$ (\ref{number}) of the kinematic space is the same as the conditional mutual information, i.e.,~$I=\mathcal{N}$. Comparing with (\ref{number}) and the expression of the conditional mutual information, one immediately has
$\Theta_{I} \propto \frac{c}{L^{D}}$,
where  $c$ is the central charge of the boundary system. This constant can be explained as the ``density'' of each isometry tensor. It counts the number of information bits (or entangled pairs) in each isometry. In Section~\ref{complexity-action}, we can see that one outstanding improvement in this paper is that we do not identify the number of the isometry as the volume, but rather the on-shell action, because (\ref{TN}) is equivalent to the number of quantum gates acting on an entangled pair in MERA. When $\Theta_{I}$ is a~constant, the on-shell action is proportional to the volume, then our statement reduces to the usual one as mentioned above.

Now we treat the (continuous) MERA as a~dS geometry rather than the AdS time slice. This~implies the emergent dimension is temporal rather than spatial. That means the opposite direction of coarse graining can be viewed as the evolution time $\tau$ of the universe. We can write down the FRW metric for this tensor network
\begin{equation}\label{metric}
ds^{2}_{\text{TN}}=-d\tau^{2}+a^{2}(\tau)d{\bf x}^{2}.
\end{equation}

For dS one has $a=\exp(\tau/L)$, where $L$ is the dS radius. One should note that our proposal can be applied to general $D$-dimensional case. Actually, since kinematic space is just an auxiliary space, it is possible to go beyond kinematic space picture to set up the connections between tensor networks and spacetime structure (\ref{metric}), and to discuss this model in cosmology.

\section{Fisher Information Measure = Gravitational Action}

As explained above, the number of isometric tensors can be given by the integral (\ref{number}). In this section we show that the integral (\ref{number}) can be regarded as the on-shell action in dS background, with the help of an observation that the on-shell action of dS spacetime can be viewed as a~FIM~\cite{Chimento:2002}.

We can even consider a~more general $D$-dimensional spacetime. For a~$D$-dimensional spacetime without matter, the total action is given by
\begin{eqnarray}
I_{\text{G}} = \frac{1}{16\pi G_{D}}\int_{\mathcal{M}} d^{D}x \sqrt{-g} (R-2\Lambda)
 + I_{GHY},\label{action}
\end{eqnarray}
where the first term is the EH action and the second is the Gibbons-Hawking-York (GHY) boundary term. $R$ is the Ricci curvature, $\Lambda=\frac{D-2}{2D}R$ is the cosmological constant and $G_{D}$ is the $D$-dimensional gravitational constant. For $D$-dimension FRW metric (\ref{metric}), the Ricci curvature is $R = 2(D-1)\frac{\ddot{a}}{a}+(D-1)(D-2)\frac{\dot{a}^2}{a^2}$.
Particularly, We consider the on-shell action (\ref{action}) for dS. The Gibbons-Hawking-York boundary terms, which are used to supplement the action so that the variational principle is well-defined, are given by the extrinsic curvature of the boundary surface $\partial\mathcal{M}$ of the spacetime $\mathcal{M}$. In general this term reads
\begin{equation}
I_{GHY} = \frac{1}{8\pi G_{D}}\int_{\partial\mathcal{M}}d^{D-1}x\epsilon \sqrt{h} K, \label{GHY}
\end{equation}
where $K$ is the trace of extrinsic curvature of the boundary and $h_{ij}=a^{2}\delta_{ij}$ is the induced metric. $\epsilon$ is equal to $+1$ ($-1$) if $\partial\mathcal{M}$ is timelike (spacelike). The boundaries of the dS spacetimes are the spacelike $\tau$ time slices at $\tau_{i}$ and $\tau_{f}$, with outward pointing unit normal $n^{\mu}$ and $n_{\mu}n^{\mu}=\epsilon=-1$. In the FRW metric we have $n^{\mu} = (1,\bf{0})$ at $\tau_{f}$ and $n^{\mu} = (-1,\bf{0})$ at $\tau_{i}$. The trace of extrinsic curvature is given by
\begin{equation}
K = \nabla_{\mu}n^{\mu} = h^{\alpha\beta}\left( \partial_{\beta}n_{\alpha}-\Gamma^{\gamma}_{\alpha\beta}n_{\gamma} \right) = (D-1)\frac{\dot{a}}{a}
\end{equation}

The GHY boundary term for the $\tau=\tau_{f}$ slice is obtained
\begin{equation}
I^{(f)}_{GHY}=-\frac{D-1}{8\pi G_{D}}\int_{\partial\mathcal{M}_{f}}d^{D-1}x a^{D-1}H,
\end{equation}
where we have used $\sqrt{h}=a^{D-1}$. Similarly, we can also obtain the contribution of $\tau=\tau_{i}$ slice $I^{i}_{GHY}$. And then we can write the total surface integral (\ref{GHY}) as a~volume integral through
\begin{eqnarray}
I_{GHY} = -\frac{D-1}{8\pi G_{D}}\int_{\mathcal{M}} d^{D-1}xd\tau \frac{d}{d\tau}\left( a^{D-2}\dot{a} \right)
\end{eqnarray}

For the dS case, this boundary term reads
\begin{equation}
I_{GHY} = -\frac{(D-1)^2}{8\pi G_{D}}\int_{\mathcal{M}} d^{D}x a^{D-1}H^{2},
\end{equation}
where $d^{D}x\equiv d^{D-1}xd\tau$.

After taking the GHY boundary term into consideration, the whole on-shell action reduces to
\begin{equation}
I_{G
} = -\frac{(D-1)(D-2)V_{D-1}}{8\pi G_{D}}\int d\tau a^{D-1}\left( \frac{\dot{a}}{a}\right)^2. \label{FRW action}
\end{equation}
where $V_{D-1}$ is the $(D-1)$-dimensional comoving volume. One of the main results in this paper, as~will see below, is to notice that this form of on-shell action can be regarded as FIM of gravity system, and that it exhibits the same behavior as the complexity by assuming reliability of the CA duality for~dS. 

Now let us turn to see how the gravitational action (\ref{FRW action}) matches a~Fisher information measure (FIM). A FIM is a~measure of the information or the disorder of a~system and has been studied in estimation theory for many years~\cite{Frieden:1998,Beck:1993}. Consider a~system specified by a~parameter $\theta$. Let $y$ be the data~value and $x$ be the noise value, we have $y=\theta+x$. There is a~function to estimate the parameter $\hat{\theta}(y)$ from data~$y$. The question is how well $\theta$ can be estimated. The answer is related to the fluctuation of data~value $y$ which can be described by a~probability density function (PDF) $p(y|\theta)$. If the translation invariance holds: $p(y|\theta)=p(y-\theta)=p(x)$, i.e.,~$p$ is only the description of noise. Then the FIM is of the following definition~\cite{Frieden:1998}
\begin{equation}
I_{\text{FIM}}\left[p\right] = \int dx\left( \frac{dp(x)}{d x} \right)^2\frac{1}{p(x)}.
\end{equation}

By introducing a~mean-square error $e^2=\int dy(\hat{\theta}(y)-\theta)^2p$, we always have $e^2I_{\text{FIM}}\ge 1$~\cite{Frieden:1998} (Appendix~\ref{FIM_def} for detail). This uncertainty relation means a~well estimation (small $e^2$) leads to a~larger $I_{\text{FIM}}$. Hence $I_{\text{FIM}}$ is a~quality of the estimation
procedure and we call it ``information''.

One can also introduce a~more general PDF called ``escort probabilities'' which is defined as~\cite{Chimento:2002,Beck:1993}
\begin{equation}
P_{q}(x)=\frac{p(x)^q}{\int dx p(x)^q}=\frac{p(x)^q}{Q}, \label{escort}
\end{equation}
where $q$ is a~real parameter and $Q=\int dx p(x)^q$ . Then a~new FIM $I_q$ can be defined in a~similar way

\begin{equation}
I_q\equiv \frac{Q}{q^2}I_{\text{FIM}}\left[P_q\right]=\int dx p(x)^{q-2}\left( \frac{d p(x)}{d x} \right)^2.
\end{equation}

$I_q$ also has information significance of the system.

Comparing $I_q$ with the on-shell gravitational action (\ref{FRW action}) and setting
\begin{equation}
q=D-1, ~~~~x=\frac{8\pi G_{D}}{(D-1)(D-2)V_{D-1}}\tau, ~~~~p(x)=a(\tau),
\end{equation}

One finds the FIM has the same form as the gravitational action, i.e.,~$I_q=-I_{G}$ and $q$ is related to spacetime dimension $D$. It is well known that the positive cosmological constant solution of the vacuum Einstein equations is the dS spacetime: $a(\tau)=\text{exp}(\tau/L)$. Then (\ref{FRW action}) can be written as
\begin{equation}
I_q=-I_{G}= \frac{(D-1)(D-2)V_{D-1}}{8\pi G_{D} L^2}\Lambda_c^{D-1},\label{FIM-action}
\end{equation}
where $\Lambda_c\equiv e^{\Lambda_{f}/L}$, $\Lambda_{f}$ is a~future cutoff on $\tau$. This implies that the FIM (or the on-shell action) of a~dS spacetime is proportional to its spacetime's volume.

Although the on-shell de Sitter action has the same form as Fisher information after regarding the scale factor $a(\tau)$ as PDF $p(x)$. We must point out that $a(\tau)$ in our case is different from a~probability density in that $a(\tau)$ is exponential. So, unlike PDF, such a scale factor has a normalization issue and its integral diverges at infrared point of de Sitter spacetime. We also need to emphasize that in this article we only consider the vacuum de Sitter case. Our model is a~toy model which comes from the picture of tensor network/gravity correspondence. Such toy model relies heavily on conformal symmetry and studies beyond AdS or de Sitter case still lack. Nevertheless, one can introduce energy-momentum tensor to the right hand side of Einstein equation as a~source of matter, such as dust or radiation for our real universe. For a~perfect fluid, the scale factor is given by $a(\tau)\sim\tau^{\alpha}$, where $\alpha<1$($\alpha=2/3$ at dust domination and $\alpha=1/2$ at radiation domination). More generally, if $\alpha$ is not a~constant, the normalization issue would not occurs anymore~\cite{Chimento:2002}. However, we still lack knowledge of the correspondence between tensor network (or circuit) and gravity with general sources.

\section{Complexity Interpreted as FIM}\label{complexity-action}
Quantum complexity is the minimum number of elementary operations in producing the target state in question from a~reference state. Here we show the complexity can be interpreted as the on-shell action, or equivalently as shown above, the FIM.

Firstly, recall that the volume $\int d^2x \sqrt{-g}$ can be equivalently given by the number of isometries. Therefore $\Theta_I$ in (\ref{number}) has the meaning of ``density'' of bits, that is, the number of information bits in each isometry. There is a~similar concept called entanglement density~\cite{Nozaki:2013wia}, which counts the number of disentanglers (tensors acting on an entanglement pair) in each bond in the tensor network. The~entanglement entropy of an interval can be obtained by roughly counting the number of bonds cut by the causal cut and then multiplying by the density. However, as to tensor network associated with kinematic space, counting the number of entanglement pairs in each isometry (i.e., $\Theta_I$) is more straightforward. As an explicit example, the conditional mutual information is just given by multiplying the density $\Theta_I$ by the number of isometry $\int d^2x \sqrt{-g}$ in a~diamond $D$. Secondly, complexity, by definition, has the meaning as the minimum number of elementary gates necessary to produce a~state $|\Psi\rangle$ from a~simple reference state $|\Psi_0\rangle$. For MERA in question, the elementary gate is the gate acting on an entanglement pair\footnote{The reason we choose element gate like this is that the MERA network can be thought of as an iterative compression algorithm that maps the density matrix of an interval to a~direct product state~\cite{Czech:2015kbp}. In the opposite direction, this tensor network maps a~non-entangled state to an entangled state~\cite{Bao:2017qcc}. Then each of the element quantum gates acts on the simplest entanglement pair ($2$-qubit). The simplest toy example of gate set we choose may be shown in Appendix~\ref{Simple_gate}.}, which means the gates we choose are the simple gates that operate on a~very small number of bits~\cite{Susskind14}. If we treat IR state of MERA as reference state and UV state as target state, and recall the MERA/dS correspondence introduced in the last section, we~conclude that evolution of dS universe can be regarded as a~process of quantum circuit from one state to another~\cite{Bao:2017qcc,Bao:2017qmt}. And the complexity of MERA is naturally determined by the number of these quantum gates acting on entangled pairs, which is proportional to $\int d^2x \sqrt{-g}\Theta_I$, namely Appendix~\ref{Simple_gate},
\begin{equation}
\mathcal{C}\propto \int d^Dx\sqrt{-g}\Theta_{I} \propto \frac{c}{L^D}\int d^Dxa^{D-1}.\label{complexity}
\end{equation}

If all these hold, we have $\mathcal{C}\propto -I_{G}$, where the minus sign comes from the fact that the manifold is Lorentzian. If we turn it to an Euclidean one by $\tau\to i\tau$, $L\to iL$ the minus sign vanishes. This implies the complexity of  MERA circuit is nothing but the FIM of the spacetime, i.e.,
\begin{equation}
\mathcal{C}=\frac{1}{\pi}I_q,\label{CA}
\end{equation}
where we have associated a~prefactor in this equality\footnote{In this paper we set $c=\hbar=1$. In general, one should associate a~prefactor $\lambda/\pi$ where $\lambda$ is a~positive number. The~undetermined prefactor $\lambda$ is only determined by the choice of gate set and Hamiltonian locality. We have already claimed above that for MERA the chosen gates are simple. Hence for a~system whose Hamiltonian locality equal to quantum-gate locality, we expect $\lambda=1$ Appendix~\ref{Simple_gate}. The prefactor is precisely the same as the one obtained for the AdS black holes~\cite{Brown:2015bva,Brown:2015lvg}.}. One minor comment is the following: since dS on-shell action is proportional to spacetime volume, one cannot differentiate the CA duality from the CV. An argument of the duality between MERA circuit complexity and $D$-dimensional dS action was discussed in~\cite{Bao:2017qmt}.


One support of the duality (\ref{CA}) comes from a~classical relation between central charge of the boundary theory and the gravitational constant of the gravity theory. From (\ref{FIM-action}) and (\ref{complexity}) we have $c\sim\frac{L^{D-2}}{G_D},$
which coincides with the well-known relation in the AdS$_3$~\cite{Brown:1986nw}. This implies for fixed $L$, the~large $c$ limit of the boundary theory will lead to a~classical gravity.

One more evidence of this relation refers to Lloyd's conjecture~\cite{Lloyd2000nat}. This conjecture claims that, if a~set of orthogonal gates $G_{\alpha}$ are chosen to construct a~target state, there is a~lower bond for the computational time that takes a~quantum system to reach an orthogonal state. This implies an upper bond for the growth rate of complexity
\begin{equation}
\frac{d\mathcal{C}}{d\tau}\le\frac{2E}{\pi},\label{bond}
\end{equation}
where $E$ is the energy of this system.
After associating the prefactor in relation of complexity and action, $\mathcal{C}=I/\pi$~\cite{Brown:2015bva,Brown:2015lvg}, the authors proposed that in the bulk the black holes obey this upper bond for the growth rate of complexity, i.e., $d\mathcal{C}/d\tau=2M/\pi$, where $M$ is the mass of an uncharged black hole. However, they use the simple gates which only operate a~small number of bits rather than the orthogonal gates~\cite{Susskind14,Cottrell:2017ayj}.

As to our model, considering (\ref{FRW action}) and (\ref{CA}), the growth rate of complexity is given by
\begin{eqnarray}
\frac{d\mathcal{C}}{d\tau}&=&\frac{(D-1)(D-2)}{8\pi^2 G_{D}}\int d^{D-1}x\sqrt{h}~\frac{\dot{a}^2}{a^2}=\frac{2E_{u}}{\pi},
\end{eqnarray}
where we have used the Friedmann equation $(D-1)(D-2)H^2=16\pi G_{D}\rho$ and $\sqrt{h}=a^{D-1}$ is the determinant of spatial component of FRW metric. Note that $E_{u}=\int d^{D-1}x\sqrt{h}\rho$ is the energy of this universe. In other words, the energy of dS universe plays the role of the energy in Lloyd's bond (\ref{bond}), just like the black hole's mass in AdS spacetime. We find that if we write the on-shell action of gravity as the form of the FIM (\ref{FRW action}), the growth rate of complexity saturates the Lloyd's bond.

\section{Complexity Interpreted as $L^n$ Norm FIM}
The proposal of interpreting complexity as FIM can be also tested in theories beyond the Einstein's gravity. In this section we try to provide more evidences to support this proposal. We firstly connect the dS on-shell action in $f(R)$ gravity to $L^n$ norm FIM and then argue the equivalence of these FIM and MERA complexity defined by Fubini-Study metric.
\subsection{Testing in $f(R)$ Gravity}
In this subsection we consider the complexity of the gravitational theory beyond the standard Einstein's gravity with certain different Ricci curvature term, the $f(R)$ gravity.

The growth rate of complexity of $f(R)$ gravity for AdS black holes has been considered in~\cite{Alishahiha:2017hwg} and it also saturates the complexity growth rate bound. We assume the prefactor between action and complexity is the same as the AdS black holes case~\cite{Brown:2015bva,Brown:2015lvg}, i.e., $\lambda=1$ for simplicity. The corresponding gravitational action of $f(R)$ gravity is given by bulk term, GHY boundary term and matter term:
\begin{eqnarray}
I_{G} = \frac{1}{16\pi G_{D}}\int_{\mathcal{M}}d^{D}x\sqrt{-g}f(R)
+ \frac{1}{8\pi G_{D}}\int_{\partial\mathcal{M}}d^{D-1}x\epsilon\sqrt{h}f'(R)K + I_{M},
\end{eqnarray}
where $f'(R)\equiv df(R)/dR$. From this action the equations of motion is derived as
\begin{eqnarray}
f'(R)R_{\mu\nu}-\frac{1}{2}f(R)g_{\mu\nu}+\left( g_{\mu\nu}\square-\nabla_{\mu}\nabla_{\nu} \right) f(R)=8\pi G_{D}T^{(M)}_{\mu\nu},
\end{eqnarray}
where $T^{(M)}_{\mu\nu}$ is the stress tensor corresponding to the matter contribution $I_{M}$. This modified field equation can be written as the standard form of the Einstein's gravity~\cite{Capozziello:2002rd}, i.e.,
\begin{equation}\label{f(R)_e.o.m}
R_{\mu\nu}-\frac{1}{2}Rg_{\mu\nu} = 8\pi G_{D}\left( \tilde{T}^{(curv)}_{\mu\nu}+\tilde{T}^{(M)}_{\mu\nu} \right),
\end{equation}

There are two contributions of the stress tensor, one comes from the matter and the other comes from the curvature. These effective stress tensors are given by
\begin{eqnarray}
\tilde{T}^{(M)}_{\mu\nu}=\frac{T^{(M)}_{\mu\nu}}{f'(R)},~~~~~~~~
\tilde{T}^{(curv)}_{\mu\nu}=\frac{1}{8\pi G_{D}f'(R)}\left\{ \frac{g_{\mu\nu}}{2}\left[ f(R)-Rf'(R) \right]-\left( g_{\mu\nu}\square-\nabla_{\mu}\nabla_{\nu} \right)f(R) \right\}.
\end{eqnarray}

We note that the effective stress tensor associated with the matter should be modified by a~factor $1/f'(R)$. Here we define a~new stress tensor $T^{(curv)}_{\mu\nu}$ of the curvature similar to the matter term as
\begin{equation}\label{curvature_stress}
T^{(curv)}_{\mu\nu} = \tilde{T}^{(curv)}_{\mu\nu}f'(R).
\end{equation}

We suggest that $T^{(curv)}_{\mu\nu}$,  like the matter stress tensor ${T}^{(M)}_{\mu\nu}$, contribute the energy density and pressure to the spacetime in our complexity's proposal.
If we consider the flat universe with the FRW metric, we obtain one of the modified Friedmann equations
\begin{equation}\label{f(R)_Friedmann}
H^{2} = \frac{16\pi G_{D}}{(D-1)(D-2)f'(R)}\left( \rho^{(curv)}+\rho^{(M)} \right),
\end{equation}
where the energy density are $\rho^{(curv)}=T^{(curv)}_{00}$ and $\rho^{(M)}=T^{(M)}_{00}$, respectively.

In general it's hard to solve the Equation \eqref{f(R)_e.o.m}. However here we only look for the solution included the dS solution we are most interested in. This solution satisfies $R_{\mu\nu}=(D-1)H^{2}g_{\mu\nu}$. Then the trace of equation of motion (\ref{f(R)_e.o.m}) is obtained
\begin{equation}\label{f(R)_e.o.m_trace}
Rf'(R) = \frac{D}{2}f(R).
\end{equation}

Note that now there is no matter term $T^{(M)}_{\mu\nu}$=0 and the energy density of curvature is $\rho^{(curv)}=-[f(R)-Rf'(R)]/2$.

To evaluate the gravitational action $I_{G}$ we should also consider the GHY boundary terms. As~before, we let these boundary surfaces be the spacelike surface ($\epsilon=-1$) at the beginning and end of the spacetime. After taking into account these surfaces we can still write the surface integral as volume integral
\begin{equation}
I_{GHY} = -\frac{(D-1)^{2}}{8\pi G_{D}}\int_{\mathcal{M}}d^{D}xa^{D-1}H^{2}f'(R).
\end{equation}

The total gravitational action is obtained
\begin{eqnarray}
I_{G} = -\frac{(D-1)(D-2)}{8\pi G_{D}}\int_{\mathcal{M}}d^{D}x\sqrt{-g} H^{2}f'(R),\label{f(R)_action}
\end{eqnarray}
where we have used $\sqrt{-g}=\sqrt{h}=a^{D-1}$ and the equation of motion (\ref{f(R)_e.o.m_trace}). The growth rate of complexity of this case reads
\begin{eqnarray}
\frac{d\mathcal{C}}{d\tau}=-\frac{1}{\pi}\frac{dI_{G}}{d\tau}&=&\frac{(D-1)(D-2)}{8\pi^2 G_{D}}\int_{\partial\mathcal{M}}d^{D-1}x\sqrt{h}H^{2}f'(R) \nonumber\\
&=& \frac{2E^{(curv)}}{\pi},
\end{eqnarray}
where we have used the Friedmann Equation (\ref{f(R)_Friedmann}). $E^{(curv)}$ is the energy from the contribution of stress tensor of curvature. It's interesting enough to see that for the $f(R)$ gravity the complexity growth rate is also bounded by the Lloyd's bound. Note that to find this relation we have used the definition of energy density $\rho^{(curv)}$ from $T^{(curv)}_{\mu\nu}$ (\ref{curvature_stress}).

Now we interpret such action in Fisher information theory. From equation of motion (\ref{f(R)_Friedmann}) we know the solution included dS is given by $f(R)\sim R^{D/2}$, which is the higher order term of curvature. The on-shell action of $f(R)$ de Sitter then is given by
\begin{equation}
I_{G}=-\frac{(D-1)(D-2)V_{D-1}}{8\pi G_{D}}\int d\tau a^{D-1}\left( \frac{\dot{a}}{a} \right)^{D}
\end{equation}

We find that in the Einstein's gravity, the Einstein-Hilbert action involving $R$, the first-order of curvature, is corresponding to the Fisher information measuring the second-order error $e^2$. While considering the $f(R)$ gravity which has the higher order term, such as $R^2$ term for our universe, this gravitational action can be regarded as the Fisher information measuring the error $e^{4/3}$, and so on. It~looks natural because in dS universe we have $R\sim H^2\sim(\dot{a}/a)^2$. The higher-order action will give us higher order $\dot{a}$, which results in the different order error from the H$\ddot{\text{o}}$lder's inequality \cite{Grinshpan:2010}(see Table~\ref{table}). That is, in general
\begin{equation}
\left( \int dy \left( \frac{\partial p}{\partial\theta}\right)^n\frac{1}{p^{n/m}} \right) \left( \int dy(\hat{\theta}-\theta)^mp \right) \ge 1.
\end{equation}
where $1/n+1/m=1$. And we can define the Fisher information according to $[I_{\text{FIM}}^{(n)}]^{n}e^{m}\ge 1$, that is
\begin{equation}
I_{\text{FIM}}^{(n)} = \sqrt[n]{\int dy \left( \frac{\partial p}{\partial\theta}\right)^n\frac{1}{p^{n-1}}} = \sqrt[n]{\int dx \left( \frac{dp}{dx}\right)^n\frac{1}{p^{n-1}}}
\end{equation}

Replacing the PDF by the escort probabilities (\ref{escort}) one can obtain a~new Fisher information $I_{q}^{(n)}$
\begin{equation}\label{Ln_Fisher}
I_{q}^{(n)} \equiv \frac{Q^{\frac{1}{n}}}{q}I_{\text{FIM}}^{(n)}[P_{q}] = \sqrt[n]{\int dx \left( \frac{dp}{dx} \right)^{n}p^{q-n}}.
\end{equation}

This is the $L^{n}$ norm Fisher information measure. We will calculate this FIM in next section and compare it with the complexity defined from the Fubini-Study metric. 

\begin{table}\label{table}
\includegraphics[scale=1]{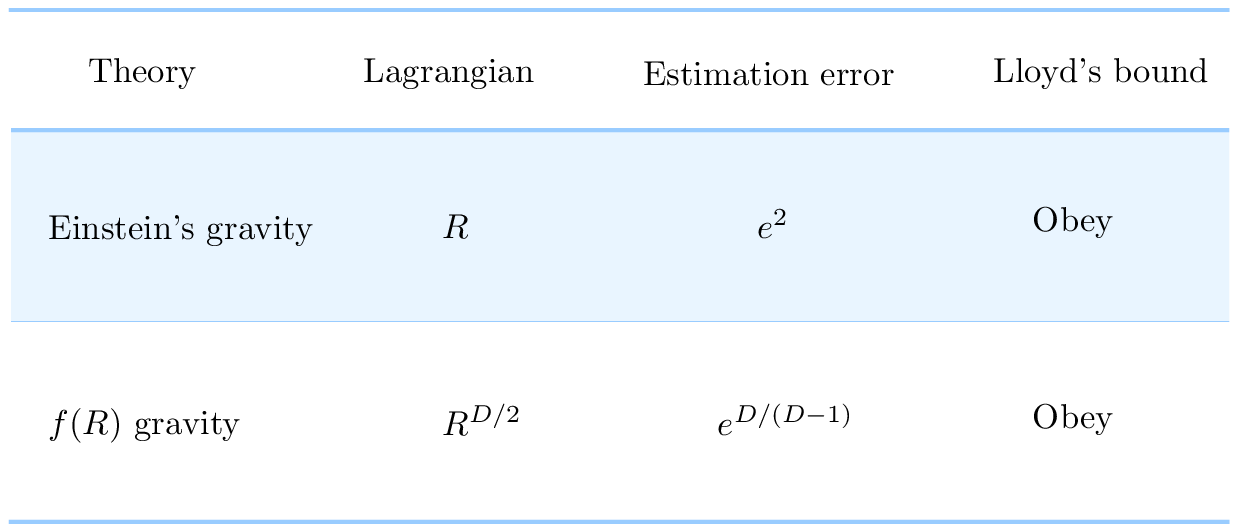}
\caption{\footnotesize{Different theories of gravity exhibit different estimation errors, but the Lloyd's bound always hold.}}\label{table}
\end{table}


\subsection{FIM as Candidates of $L^n$ Complexity}\label{FS}
Recently the definition of complexity of a~state in quantum field theory has been proposed in~\cite{Chapman:2017rqy}. We first review this proposal in the cMERA we interested in and then compare it with our definition of Fisher information measure for the dS spacetime.

The complexity $\mathcal{C}_{\text{FS}}$ in~\cite{Chapman:2017rqy} is defined by the minimal length according to the so-called Fubini-Study metric of a~path from a~referenced state $|\Psi(s_{i})\rangle=|\Psi_{0}\rangle$ to a~target state $|\Psi(s_{f})\rangle=|\Psi\rangle$. We take iterating generators $G(s)$ from some elementary set $\mathcal{G}$ and consider the unitary operators $U$ arising from G(s):
\begin{equation}
U(\sigma)=\mathcal{P}e^{-i\int_{s_{i}}^{\sigma}G(s)ds},
\end{equation}
where $\sigma\in[s_{i},s_{f}]$. Note that in $s_{f}$ we should introduce a~UV cutoff $\Lambda$. For the intermediate states $|\Psi(\sigma)\rangle=U(\sigma)|\Psi_{0}\rangle$ the Fubini-Study line element is defined as
\begin{equation}
ds_{\text{FS}}(\sigma) = d\sigma\sqrt{\langle G^2(\sigma)\rangle-\langle G(\sigma)\rangle^2}.
\end{equation}

By restricting the allowed operators $G(s)$ this distant is more non-trivial and the complexity of $|\Psi\rangle$ under these allowed operators is defined  as the minimal length according to FS metric of a~path from $|\Psi_{0}\rangle$ to $|\Psi\rangle$:
\begin{equation}\label{FS_compelxity}
\mathcal{C}_{\text{FS}}(|\Psi_{0}\rangle, |\Psi\rangle, \mathcal{G}, \lambda) = \min_{G(s)}\int_{s_{i}}^{s_{f}}ds_{\text{FS}}(\sigma).
\end{equation}

We are interested in the massless free quantum fields which is also conformal because the metric $g_{\tau\tau}$ of such case is the same as AdS or dS geometry~\cite{Nozaki:2012zj}. The corresponding tensor network is cMERA. We can calculate the complexity of the cMERA network corresponding to the Gaussian states in this quantum field by using the elementary set $\mathcal{G}=\text{Span}(K(\vec{k}))$, where $K(\vec{k})$ is the two mode squeezing operator (dis)entangles the $\vec{k}$ and $-\vec{k}$ modes. The cMERA circuit maps the Gaussian reference state $|R(M)\rangle$ which has no spatial correlations to a~approximate ground state $|m^{(\Lambda)}\rangle$:
\begin{equation}
|m^{(\Lambda)}\rangle = \mathcal{P}e^{-\frac{i}{2}\int_{-\infty}^{0}du\int_{k\le\Lambda~e^{u}}d^{D-1}kK(\vec{k})\chi(u)}|R(M)\rangle,
\end{equation}
where $\chi(u)=[e^{2u}/(e^{2u}+m^{2}/\Lambda^{2})]/2$ and $M=\sqrt{\Lambda^{2}+m^{2}}$. $u$ is a~renormalization group scale parameter from IR $u=-\infty$ to UV $u=0$, which corresponds to $\sigma\in[s_{i}, s_{f}]$. The operator $G(u)$ is given by $\int_{k\le\Lambda~e^{u}}d^{D-1}kK(\vec{k})\chi(u)/2$. From Fubini-Study distant (\ref{FS_compelxity}) we have
\begin{equation}\label{MERA_complexity}
\mathcal{C}^{(2)}_{\text{cMERA}} = \int_{-\infty}^{0}du\chi(u)\sqrt{\frac{V_{D-1}}{2}\int_{k\le\Lambda~e^{u}}d^{D-1}k}
\end{equation}
where $V_{D-1}$ is the volume of $(D-1)$-dimensional time slice in quantum field. The superscript $(2)$ implies (\ref{MERA_complexity}) is an $L^{2}$ norm. If we restrict that $\mathcal{G}$ contain only $K(\vec{k})$ and not their linear span. This leads to a~$L^{1}$ norm complexity
\begin{equation}
\mathcal{C}^{(1)}_{\text{cMERA}} = \frac{V_{D-1}}{2}\int_{-\infty}^{0}du\chi(u)\int_{k\le\Lambda~e^{u}}d^{D-1}k.
\end{equation}

One can define a~general $L^{n}$ norm as a~measure of complexity and in the massless free CFT it has~form
\begin{equation}\label{Ln_complexity}
\mathcal{C}^{(n)}_{\text{cMERA}} = \int_{-\infty}^{0}du\sqrt[n]{\frac{V_{D-1}}{2}\int_{k\le\Lambda~e^u}d^{D-1}k\left(\chi(u)\right)^{n}}
\end{equation}

In the massless case $m=0$ we have $\chi(u)=1/2$ and $M=\Lambda$, the $L^{n}$ norm complexity can be calculated  analytically
\begin{equation}
\mathcal{C}^{(n)}_{\text{cMERA}} = \frac{n}{2(D-1)}\sqrt[n]{\frac{\pi^{d/2}}{2\Gamma(\frac{D-1}{2}+1)}}V_{D-1}^{\frac{1}{n}}\Lambda^{\frac{D-1}{n}},
\end{equation}
which is proportional to $V^{1/n}_{D-1}\Lambda^{D-1/n}$. This is the only divergence term of the complexity. In general there are $\log(M/\Lambda)$ divergence terms and they vanish in the cMERA($m=0$) case because of $M=\Lambda$.

Now turn to our proposal of the $L^{n}$ norm Fisher information measure which we have obtained in (\ref{Ln_Fisher}). The $L^{n}$ norm Fisher information is a~quality metric of the estimation corresponding to $e^{n/(n-1)}$ error. After setting
\begin{equation}
q-n=D-1, ~~~~~~x=\left( \frac{8\pi G_{4}}{(D-1)(D-2)V_{D-1}} \right)^{\frac{1}{n-1}}\tau, ~~~~~~p(x)=a(\tau),
\end{equation}
we have
\begin{equation}
I_{q}^{(n)} = \sqrt[n]{\frac{(D-1)(D-2)V_{D-1}}{8\pi G_{D}}\int d\tau a^{D-1}\left( \frac{\dot{a}}{a} \right)^{n}}. \label{I_q^n}
\end{equation}
Comparing the $L^{n}$ norm Fisher information with the $L^{n}$ norm complexity of MERA, the divergence in (\ref{Ln_complexity}) comes from the integral in momentum with a~UV cutoff , which is equivalent to the divergence from the temporal integral with cutoff in $L^{n}$ norm Fisher information (\ref{I_q^n}). On the other hand, in (\ref{Ln_complexity}) $\chi(u)$ is equivalent to $(\dot{a}/a)$ in (\ref{I_q^n}). To see this, we note that the original form of the $\chi(u)$ is~\cite{Nozaki:2012zj}:
\begin{equation}
\chi(u) = \frac{1}{2}\left( \frac{|k|\partial_{|k|}\epsilon_{k}}{\epsilon_{k}} \right)\bigg{|}_{|k|=\Lambda~e^{u}},
\end{equation}
where $\epsilon_{k}=\sqrt{k^{2}+m^{2}}$. The parameter $u$ is related to dS time by $\Lambda~e^{u}=e^{\tau}$. Here we only consider the cMERA with $m=0$ and we have
\begin{eqnarray}
\chi(u) &=& \frac{1}{2}\frac{|k|\partial_{u}\sqrt{k^{2}+m^{2}}}{\sqrt{k^{2}+m^{2}}}\frac{du}{d|k|}\bigg{|}_{|k|=\Lambda~e^{u}, m=0} = \frac{1}{2}\frac{d(\Lambda~e^{u})/du}{\Lambda~e^{u}}\nonumber\\
&=& \frac{1}{2}\frac{de^{\tau}/d\tau}{e^{\tau}} = \frac{1}{2}\frac{\dot{a}}{a}.
\end{eqnarray}
So the $L^{n}$ norm complexity of cMERA are coincided with the $L^{n}$ norm Fisher information.

For the cMERA the dual gravity is a~dS spacetime $a(\tau)=e^{\tau/L}$. One can transform the FRW metric to a~comformal metric $ds^2=(-dt^{2}+d{\bf x}^{2})/t^{2}$ by using $e^{\tau/L}=1/t$. Then the UV cutoff $\tau_{f}=\Lambda_{f}$ is given by $e^{\Lambda_{f}/L}=1/\epsilon=\Lambda$. The IR state correspond to $\tau_{i}=-\infty$. Then the $L^{n}$ norm Fisher information for the dS universe is obtained
\begin{equation}\label{ln_action}
I_{q}^{(n)} = \sqrt[n]{\frac{(D-2)}{8\pi G_{D}L^{n-1}}}V_{D-1}^{\frac{1}{n}}\Lambda^{\frac{D-1}{n}}.
\end{equation}

We find that the $L^{n}$ norm Fisher information are coincided with the $L^{n}$ norm complexity of cMERA up to a~factor~\cite{Chapman:2017rqy}
\begin{equation}
I_{q}^{(n)} \sim \mathcal{C}_{\text{cMERA}}^{(n)} \sim V_{D-1}^{\frac{1}{n}}\Lambda_c^{\frac{D-1}{n}},
\end{equation}
as we expected. That means they have the same structure of divergence $\sim \Lambda_c^{\frac{D-1}{n}}$ and are also proportional to $n$-th root of the volume of time slice $\sim V_{D-1}^{1/n}$. The complexity can be regarded as the Fisher information measure corresponding to errors.

When $n=D$ these $L^{D}$ norm FIM represent the on-shell gravitational action of the $D$-dimensional $f(R)$ gravity we discussed above, i.e.,~$[\mathcal{C}_{\text{cMERA}}^{(D)}]^{D}\sim[I_{\text{FIM}}^{(D)}]^{D}\sim-I_{G}$. This receives one more support of our statement. Moreover, this similarity suggests that the $L^{n}$ norm FIM is a~candidate of the dual theory of the $L^n$ norm complexity of cMERA.

\section{Conclusions and Discussion}
In summary, based on two premises that the information interpretation of spacetime and the CA correspondence holds for dS universe, we show that isometry in MERA can be viewed as a~quantum gate which operates information bits and the dS universe may be understood as a~MERA tensor network.
More specifically, the complexity of MERA which counts the number of operations on information bits is given by the on-shell action of the dS spacetime.
On the other hand,the on-shell action can be regarded as the FIM of the ``probability density
function'' $a(\tau)$~\cite{Chimento:2002}. Therefore, in this article we further show that: (i)complexity of a~MERA network admits a~novel explanation as FIM  of dS universe, i.e., $\mathcal{C}=I_q/\pi$. We consider the total dS action including cosmological constant and GHY boundary terms, which are lacking in previous researches. (ii)We extend this statement to theories beyond Einstein's gravity. In particular, we find that dS action of $D$-dimensional $f(R)$ gravity can also be regarded as the FIM. Because $f(R)$ gravity contains higher order curvature, the corresponding FIM is the $L^{n}$ norm FIM, which measures different order error $e^{D/(D-1)}$. It turns out this $L^{n}$ norm FIM of $f(R)$ dS is a~candidate of the dual theory of the recent proposed $L^n$ norm complexity of cMERA in~\cite{Chapman:2017rqy}, where the dual theory of the proposed $L^n$ norm complexity is missing. (iii)The FIM's form of complexity(or dS action equivalently) saturates the Lloyd's bond both for Einstein's gravity and $f(R)$ gravity. In Einstein's gravity the energy of dS universe $E=\int d^{D-1}x\sqrt{h}\rho$  plays the role of the energy in Lloyd's bond, while in $f(R)$ gravity the energy in Lloyd's bond is given by the energy of curvature $E=\int d^{D-1}x\sqrt{h}\rho^{(curv)}$.

\section*{\bf Acknowledgements}
This work was supported in part by the National Natural Science Foundation of China under Grant Nos. 11975116, 11665016 and 11563006, and Jiangxi Science Foundation for Distinguished Young Scientists under Grant No. 20192BCB23007.

\appendix
\section{Fisher Information Measure: A Brief Review}\label{FIM_def}
In this section we review the derivation of the Fisher information measure from estimation theory. One can find the details for these definitions in~\cite{Frieden:1998}.

Fisher information meausre (FIM) is a~measure of how well we can estimate a~parameter $\theta$ of a~given system. Given a~series of data~$y=\theta+x$, this parameter can be estimated by an optimal function $\hat{\theta}(y)$. The system is specified by a~distribution function $p(y|\theta)$, which is called the probability density function (PDF) of the data~$y$. We expect that overall measurement procedure is well on average, i.e., $\langle \hat{\theta}(y) \rangle=\theta$. So we introduce a~mean-square error
\begin{equation}
e^2 \equiv \int dy(\hat{\theta}-\theta)^2p(y|\theta)
\end{equation}
in the estimation. The smaller $e^2$ represents the better expected estimation.

Consider the estimator $\hat{\theta}(y)$ obeying
\begin{equation}
\langle \hat{\theta}-\theta~\rangle = \int dy(\hat{\theta}-\theta)p(y|\theta)=0.
\end{equation}

PDF $p(y|\theta)$ describes the fluctuation of data~$y$ in the presence of the parameter value $\theta$. After~differentiating this equation w.r.t. $\theta$ in both sides we have
\begin{equation}\label{estimation}
\int dy(\hat{\theta}-\theta)\frac{\partial p}{\partial\theta} = \int dy p.
\end{equation}

Because of the normalization of the PDF the r.h.s is equal to $1$ and we could write this integral as
\begin{equation}
\int dy \left( \frac{\partial p}{\partial\theta}\frac{1}{\sqrt{p}} \right)\left( (\hat{\theta}-\theta)\sqrt{p} \right)=1.
\end{equation}

Squaring both sides of the equation and using the Schwarz inequality~\cite{Aldaz:2015}, we obtain
\begin{equation}
\left( \int dy\left( \frac{\partial p}{\partial\theta} \right)^2\frac{1}{p} \right) \left( \int dy( \hat{\theta}-\theta)^2p \right)\ge 1.
\end{equation}

It's obvious that the right-most term is the mean-square error $e^2$. The left-most term is defined as the FIM $I_{\text{FIM}}$. Hence we always have $e^2I_{\text{FIM}}\ge1$. It is an intrinsic uncertainty due to the outside sources of noise, which implies $I_{\text{FIM}}$ is a~quality metric of the estimation procedure.

\section{Remarks on the Quantum Circuit}\label{Simple_gate}
We make some remarks on the elementary gates in the quantum circuit of our toy model. The~prefactor of the Lloyd's bound depends on the choice of gate set $G_{\alpha}$. Hamiltonian locality implies each term $H_{\alpha}$ in the Hamiltonian is $k$-local, which means they only have a~size less than or equal to a~small $k$\footnote{In general for a~system with $N$ degrees of freedom, the concept of scrambling time that describes how long a~$O(1)$ perturbation spreads over $O(N)$ d.o.f. This concept is only valid for systems with $k\ll N$~\cite{Brown:2015lvg}.}. The gate set is suggested to be chosen at approximate unitary evolution $U(\tau)=e^{-i\delta~H}$ in a~small $\delta$, i.e., $G_{\alpha}=e^{i\delta~H_{\alpha}}\simeq I+i\delta~H_{\alpha}$, which indicates it is close to the identity and is simple ($G_{\alpha}$ is also near $k$-local). If the Hamiltonian is $k$-local with large $k\sim N$ and the gate set is $j$-local with $j\ll k$, even for small $\delta$ the amount of gates increases so fast that violates the bond. Therefore we need to modify the prefactor as~\cite{Brown:2015lvg}
\begin{equation}
\frac{d\mathcal{C}}{d\tau}\le \frac{g(k)}{f(j)}\frac{2E}{\pi},
\end{equation}
where $g$ and $f$ capture the dependence of the Hamiltonian and gate set. If $k=j$ the prefactor $g/f$ is equal to $1$.

The elementary gates we choose obey the following two requirements: (i) they are simple and (ii) create the entanglement between qubits. This is also the proposal of the model of the quantum circuit cosmology as discussed in~\cite{Bao:2017qcc}. A simple example of gate that creates the entanglement is the following. First we need a~Hadamard gate~\cite{Eleanor}: $H=\frac{1}{\sqrt{2}}\left( |0\rangle\langle0|+|0\rangle\langle1|\right)+\frac{1}{\sqrt{2}}\left(|1\rangle\langle0|-|1\rangle\langle1| \right)$, which transforms a~single-qubit state into a~new state in this way
\begin{equation}
H|0\rangle= \frac{1}{\sqrt{2}}|0\rangle+\frac{1}{\sqrt{2}}|1\rangle,~~~~~~~~~~
H|1\rangle= \frac{1}{\sqrt{2}}|0\rangle-\frac{1}{\sqrt{2}}|1\rangle,
\end{equation}

To proceed, we also need a~controlled-NOT gate, i.e., $C_{\text{NOT}}$~\cite{Eleanor,Monroe:1995gh}: $C_{\text{NOT}}=|00\rangle\langle00|+|01\rangle\langle01|+|11\rangle\langle10|+|10\rangle\langle11|$. The controlled-NOT gate is a~unitary gate that operates on $2$-qubit. It flips the second qubit if and only if the first qubit is $|1\rangle$, i.e., when operates on a~$2$-qubit state, it results in
\begin{equation}
C_{\text{NOT}}|00\rangle=|00\rangle, ~~~~~C_{\text{NOT}}|01\rangle=|01\rangle, ~~~~~C_{\text{NOT}}|10\rangle=|11\rangle, ~~~~~C_{\text{NOT}}|11\rangle=|10\rangle.
\end{equation}

The importance of the controlled-NOT gate is the ability to entangle two bits and produce a~Bell state. That is, when we operate $C_{N}$ on $1/\sqrt{2}(|0\rangle+|1\rangle)$ and $|0\rangle$, we have
\begin{equation}\label{controlled-not}
C_{\text{NOT}}\left((\frac{1}{\sqrt{2}}|0\rangle+\frac{1}{\sqrt{2}}|1\rangle)\otimes|0\rangle \right)=\frac{1}{\sqrt{2}}|00\rangle+\frac{1}{\sqrt{2}}|11\rangle
\end{equation}

\begin{figure}[H]
\centering
\includegraphics [scale=0.8]{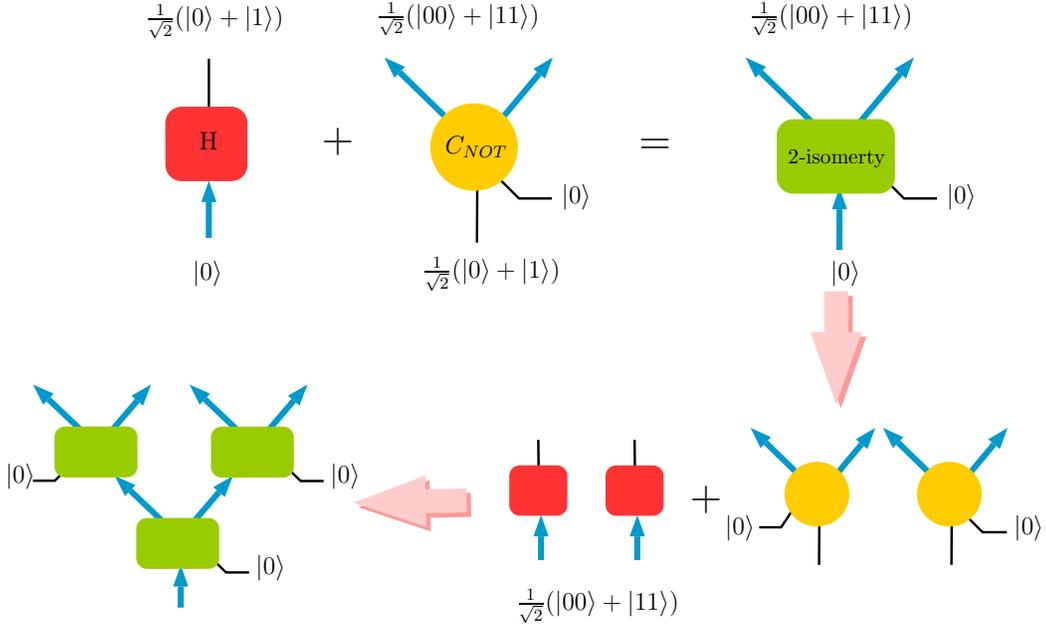}
\caption{The computation process of the MERA circuit with the Bell gates.}\label{Gate}
\end{figure}

Now it's easy to construct the elementary gate set $G_{\alpha}$ operate on $2$-qubit in our MERA circuit as
\begin{equation}
G_{\alpha}:=C_{\text{NOT}}\left( H\otimes I \right),
\end{equation}
which produces an entangled pair. The idea~of entanglement equals to geometry was proposed in~\cite{VanRaamsdonk:2010pw}. If we treat IR state of MERA as reference state and UV state as target state and recall the MERA/dS correspondence introduced in the last section, we conclude that evolution of dS universe can be regarded as a~process of quantum circuit from a~trivial state $|\Psi_{0}\rangle$ to another nontrivial entangled state $|\Psi\rangle$.
First we operate an elementary gate on $|0\rangle$ to create an entanglement pair (\ref{controlled-not}) in $\Delta~t=1$ time. This is the beginning of emerge gravity. To further entangle with other qubits , we operate two elementary gates in $\Delta~t=1$ time, one on the first qubit of $1/\sqrt{2}(|00\rangle+|11\rangle)$ and a~new qubit, and the other one on the second qubit of $1/\sqrt{2}(|00\rangle+|11\rangle)$ and another new qubit (see Figure~\ref{Gate}). And so on, at time $T$ the number of gates we need is
\begin{equation}
\mathcal{C}\sim c\sum_{t=0}^{T}2^{t}\Delta~t \sim c\sum_{t=0}^{T}e^{t}\Delta~t,
\end{equation}
where $c$ comes from the number of entanglement pairs in each isometry. This is the discrete version of the dS action $I_{G}\sim\frac{1}{G_{D}}\int  e^{\tau/L}d\tau$. In other words, for obtaining emerge gravity now day the complexity we need behaves like the on-shell action.

\end{document}